\documentclass[letterpaper,english,aps,prb,floatfix,showpacs,amsfonts,amssymb,superscriptaddress,twocolumn]{revtex4}
\usepackage[T1]{fontenc}
\usepackage[latin1]{inputenc}
\usepackage{graphics}
\usepackage{amssymb}
\usepackage{amsmath}
\usepackage{overpic}

\newcommand{\be}{\begin{equation}}
\newcommand{\ee}{\end{equation}}
\newcommand{\bea}{\begin{eqnarray}}
\newcommand{\eea}{\end{eqnarray}}

\def\e{\varepsilon}
\def\d{\delta}

\def\l{\lambda}

\def\ra{\rightarrow}

\def\lb{\label}
\def\pref#1{(\ref{#1})}

\newcount\bozza \bozza=0
\ifnum\bozza=1
\newdimen\shift \shift=-2truecm
\def\lb#1{%
{\label{#1}\rlap{\kern\shift{$\scriptstyle#1$}}}}
\else\def\lb#1{\label{#1}} \fi

\begin{document}

\title{Role of the vortex-core energy on the Beresinkii-Kosterlitz-Thouless\\
transition in thin films of NbN}

\author{Mintu Mondal} 
\affiliation{Tata Institute of Fundamental Research, Homi Bhabha Rd., Colaba, Mumbai,
  400005, India}
\author{Sanjeev Kumar}
\affiliation{Tata Institute of Fundamental Research, Homi Bhabha Rd., Colaba, Mumbai,
  400005, India}
\author{Madhavi Chand}
\affiliation{Tata Institute of Fundamental Research, Homi Bhabha Rd., Colaba, Mumbai,
  400005, India}
\author{Anand Kamlapure}
\affiliation{Tata Institute of Fundamental Research, Homi Bhabha Rd., Colaba, Mumbai,
  400005, India}
\author{Garima Saraswat}
\affiliation{Tata Institute of Fundamental Research, Homi Bhabha Rd., Colaba, Mumbai,
  400005, India}
\author{G.~Seibold}
\affiliation{Institut F\"ur Physik, BTU Cottbus, PBox 101344, 03013
  Cottbus,Germany}
\author{L.~Benfatto} 
\affiliation{ISC-CNR and Dep. of Physics, Sapienza University of Rome,
  P.le A. Moro 5, 00185, Rome, Italy}
\author{Pratap Raychaudhuri}
\affiliation{Tata Institute of Fundamental Research, Homi Bhabha Rd., Colaba, Mumbai,
  400005, India}


\date{\today}

\begin{abstract}
  We analyze the occurrence of the Beresinkii-Kosterlitz-Thouless
  transition in thin films of NbN at various film thickness, by probing
  the effect of vortex fluctuations on the temperature dependence of the
  superfluid density below $T_{BKT}$ and of the resistivity above
  $T_{BKT}$. By direct comparison between the experimental data and the
  theory we show the crucial role played by the vortex-core energy in
  determining the characteristic signatures of the BKT physics, and we
  estimate its  dependence on the disorder level. Our work provides a
  paradigmatic example of BKT physics in a quasi-two-dimensional
  superconductor.
\end{abstract}


\maketitle

Ever since the pioneering work of Berezinskii, Kosterlitz and Thouless
(BKT)\cite{bkt,bkt2} 
predicting the occurrence of a phase transition without a continuously broken symmetry
in quasi 2 dimensional (2D) systems, a lot of effort has been devoted to study
its realization in real materials\cite{review_minnaghen}. Of particular interest have
been 2D superconductors\cite{Matthey,hebard,Reyren,armitage_cm10,kamlapure_apl10,triscone_nature08,iwasa_natmat10,budhani}, 
where the superconducting transition is expected to belong to
the BKT universality class. In these systems, the BKT transition can be
studied through two different schemes. When approaching the transition
temperature $T_{BKT}$ from below, the superfluid density ($n_s$) (which is related to the magnetic
penetration depth $\l$) is expected to go to zero discontinuously at the
transition, with an ``universal'' relation between $n_s(T_{BKT})$ and $T_{BKT}$
itself\cite{nelson_prl77,review_minnaghen}. Approaching the transition from above, one can identify the BKT
transition from superconducting fluctuations, which leave their signature in
the temperature dependence of various quantities, such as resistivity or
magnetization\cite{HN}. In this second scheme, the information on the BKT transition
is encoded in the correlation length $\xi(T)$, which diverges exponentially at
$T_{BKT}$, in contrast to the power-law dependence expected within Ginzburg-Landau
theory\cite{varlamov_book}.  

Many of the experimental investigations on 2D superconductors have relied
on this second approach\cite{Matthey,Reyren,triscone_nature08,iwasa_natmat10} to identify the BKT transition through the temperature dependence of resistivity $\rho (T)$, using eventually the interpolating formula proposed long ago
by Halperin and Nelson\cite{HN} to describe the crossover from BKT to
ordinary GL superconducting fluctuations. However, real superconductors
have additional complicacies that can make such an analysis more involved
than what has been discussed in the original theoretical approach. 
First, real systems always have some degree of inhomogeneity,
which tends to smear the sharp signatures of BKT transition compared to the
clean case. As it has been recently shown through scanning tunneling
spectroscopy measurements\cite{sacepe09,mondal_prl11}, even when disorder in
the system is homogeneous, the system shows intrinsic tendency towards the
formation of spatial inhomogeneity in the superconducting state, which has
to be taken into account while analyzing the BKT transition. At a more
fundamental level, it has recently been proved experimentally\cite{kamlapure_apl10}
that for a real superconductor the vortex core energy ($\mu$) can be very different from the
value predicted within the 2D $XY$ model, that was originally investigated by
Kosterlitz and Thouless as the paradigmatic case to study the BKT
transition\cite{bkt2}. This can give rise to a somehow 
different manifestation of the vortex physics, even without the extreme of
a change of the order of the transition, as it has been proposed in the past\cite{review_minnaghen,gabay}. 
Recently the relevance of $\mu$ for the
BKT transition has attracted a renewed theoretical interest in different context,
ranging from the case of layered high-temperature
superconductors\cite{benfatto_mu,benfatto_bilayer,podolsky} to the one of
superconducting interfaces in artificial
heterostructures\cite{benfatto_inho}.

All the above issues explain why more than 30 years after the prediction of
the BKT transition in ultrathin films of superconductors its occurrence in
real materials is still controversial. The present work aims to give a
paradigmatic example of the emergence of the BKT transition in thin films
of NbN as the film thickness decreases.  By a systematic comparison between
$\l (T)$ and $\rho (T)$, we show that to fully
capture the `conventional' BKT behavior in a real system one must account
for the correct value of $\mu$ as compared to the energy
scale given by the superfluid stiffness $J_s$. The analysis carried out for films
of different thickness provides us also with an indirect measurement of the
dependence of the vortex-core energy on disorder, showing that vortices
become energetically more expensive as disorder increases. Such a result
can be related to the separation between the energy scales connected to the
SC gap ($\Delta$) and $J_s$ as disorder increases, as we show by 
computing the ratio $\mu/J_s$ within the Bogoliubov-de-Gennes (BdG)
solution of the attractive Hubbard model with
on-site disorder. Our results shed new light on the occurrence of the BKT
transition on disordered films.

Our samples consist of epitaxial NbN films grown on single crystalline
(100) oriented MgO substrates with thickness ($d$) varying
between 3-50 nm. The deposition conditions were optimized to obtain
the highest $T_c$ (~16K) for a 50 nm thick film. Details o sample preparation and characterisation have been reported elsewhere\cite{choka_nbnpaper,kamlapure_apl10}. The absolute
value of $\l$ as a function of temperature was measured using a low-frequency
(60 kHz) two-coil mutual inductance technique\cite{kamlapure_apl10} on 8 mm diameter films
patterned using a shadow mask. $\rho(T)$ was measured on the same
films through conventional 4-probe technique after pattering the films into
1 mm$\times$6 mm stripline using Ar-ion milling.

\begin{figure}[ttt]
\includegraphics[scale=0.5,clip=]{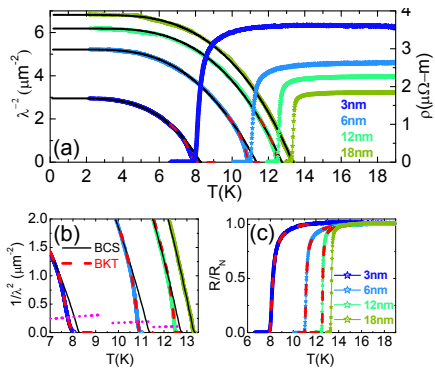}
\caption{(Color online) (a) Temperature dependence of $\lambda^{-2}(T)$ and $\rho(T)$  for
  four NbN films with different thickness. The (black) solid lines and (red)
  dashed lines correspond to the BCS and BKT fits of the $\lambda^{-2}(T)$
  data respectively. (b) An expanded view of $\lambda^{-2}(T)$ close to
  $T_{BKT}$; the intersection of the BCS curve with the dotted line is where the
  BKT jump would be expected within the $XY$ model, when $\mu$ is large. (c) Temperature variation of $R/R_N$. The (red) dashed lines show the theoretical fits
  to the data, as described in the text.}
\label{fig-rhos}
\end{figure}

The first clear signature of the presence of vortices in our samples is
provided by the deviations of $\lambda^{-2}(T)$ from the BCS temperature
dependence as $d$ decreases. In particular, we observe a
sharp downturn of $\lambda^{-2}(T)$, reminiscent of the
so-called universal jump of the superfluid density\cite{nelson_prl77}. 
 To clarify the notation, we
recall that for a 2D superconductor $J_s$ is defined as:
\be
\lb{defj}
J_s=\frac{\hbar^2 n_s^{2d}}{4m}=\frac{\hbar^2 c^2 d}{16\pi e^2 \l^2 }.
\ee
where $n_s^{2d}$ is the effective 2D superfluid density. In a conventional 3D superconductor $J_s(T)$ goes to zero
continously at the SC temperature $T_c$. Instead within BKT theory the SC transition is controlled by
the vortex-antivortex proliferation, that becomes entropically favorable at
the temperature scale $T_{BKT}$ defined self-consistently by the relation
\be
\lb{jump}
\frac{\pi J_s(T_{BKT})}{T_{BKT}} = 2.
\ee
In the above relation the temperature dependence of $J_s(T)$ is due not
only to the existence of quasiparticle excitations above the gap, but also
to the presence of bound vortex-antivortex pairs {\em below}
$T_{BKT}$. The latter effect is usually negligible when $\mu$ is large, as it is the case for superfluid films\cite{helium_films}. In this case
one can safely estimate $T_{BKT}$ as the temperature where the line
$2T/\pi$ intersects the $J_s^{BCS}(T)$ obtained by a BCS fit of the
superfluid stiffness at lower temperatures. However, as $\mu$ decreases the renormalization of $J_s$ due to bound vortex pairs
increases, and consequently $T_{BKT}$ is further reduced with respect to
$T_c$.\cite{benfatto_mu,kamlapure_apl10} To account for this effect we fitted the temperature dependence of
$\l^{-2}(T)$ by integrating numerically the
renormalization-group equations of the BKT
theory\cite{review_minnaghen,benfatto_mu}  using as only free parameter\cite{kamlapure_apl10} the ratio
$\mu/J_s$. As 
input parameter for $J_s(T)$ we used the one obtained by a BCS fit of the
data (solid lines in Fig.\ 1a) at low temperatures, where vortex excitations are suppressed, which
extrapolates to zero at the mean-field transition temperature $T_c$.  As one can see
in Fig. 1b, the transition is still slightly rounded near $T_{BKT}$, so
that the sharp jump is replaced by a rapid downturn at the intersection
with the universal $2T/\pi$ line. We attribute this effect to the spatial
inhomogeneity of the sample, that can be accounted for by assuming a 
distribution of local $J_s^i(T)$ values
around the BCS one, and performing an average of the $\lambda^{-2}(T)$ 
associated to each patch. For simplicity we assume that the occurrence probability
$w_i$ of each local $J_s^i$ value is Gaussian,  with relative width
$\delta$. We then rescale proportionally the local $T_c^i$ and we
calculate the resulting $T_{BKT}^i$ from the RG equations\cite{benfatto_bilayer,kamlapure_apl10}
As shown in Fig. 1a-b, such a procedure leads to an excellent fit of the
experimental data in the whole temperature range. The obtained values of 
the ratio $\mu/J_s$ (Table I) are relatively small  
as compared to the standard expectation of the
$XY$ model\cite{nagaosa_book}, where 
\be
\lb{muxy}
\frac{\mu_{XY}}{J_s}\simeq \frac{\pi^2}{2}\simeq 4.9.
\ee
We recall that within the BKT approach to the $XY$ model
the value of $\mu$ is fixed by the cut-off
at short length scale of the energy of a vortex line, 
\be
\lb{etot}
E=\pi J_s\left[ \log \frac{L}{\xi_0}+ \alpha\right]
\ee
where $L$ is the system size, $\xi_0$ is the coherence length and
$\mu\equiv \pi J_s\alpha$.  By mapping the (lattice) $XY$ model into
the continuum Coulomb-gas problem\cite{nagaosa_book} one obtains
$\alpha\simeq \pi/2$, so that $\mu$ attains the value in eqn. \pref{muxy}. However,
in our samples $\mu$ is better estimated from the loss of condensation
energy within the vortex core (see discussion below), leading to a smaller 
$\mu/J_s$ ratio and to the deviations of the data from the BCS fit already before the
renormalized stiffness reaches the universal value $2T/\pi$. 

To further establish the validity of the values of $\mu$ obtained
from the behavior of $\l^{-2}(T)$ {\em
  below} $T_{BKT}$ we now use the same set of parameters to analyze the $\rho(T)$ {\em above}
$T_{BKT}$. In 2D, the contribution of SC fluctuations to the
conductivity can be encoded in the temperature dependence of the SC
correlation length, $\delta\sigma\propto \xi^2(T)$. The functional
form of $\xi(T)$ depends on the character of the SC fluctuations, being
power-law for Gaussian Ginzburg-Landau (GL) fluctuations\cite{varlamov_book} and
exponential for BKT-like vortex fluctuations\cite{bkt2,HN}.  
Due to the proximity
between $T_{BKT}$ and $T_c$ (Table I), we expect that most of the fluctuation
regime for the paraconductivity will be described by standard
GL SC fluctuations, while vortex fluctuations will be relevant
only between $T_c$ and $T_{BKT}$. To interpolate between the two regimes we
resort then to the Halperin-Nelson formula for $\xi$\cite{HN}
\be
\lb{corr}
\frac{\xi}{\xi_0}=\frac{2}{A}\sinh \frac{b}{\sqrt{t}}
\ee
where $t=(T-T_{BKT})/T_{BKT}$ and $A$ is a constant of order one. 
$b$ is the most relevant parameter to determine the shape of the
resistivity above the transition and is connected\cite{benfatto_inho} both to the
relative distance $t_c$ between $T_{BKT}$ and $T_c$, 
$t_c=(T_c-T_{BKT})/T_{BKT}$, and to the value of $\mu$: 
\be
\lb{btheo}
b_{theo}\sim \frac{4}{\pi^2}\frac{\mu}{J_s}\sqrt{t_c}
\ee
The normalized resistance corresponding to the SC correlation length \pref{corr} is
given by
\be
\lb{res}
\frac{R}{R_N}=\frac{1}{1+(\Delta\sigma/\sigma_N)}\equiv
\frac{1}{1+(\xi/\xi_0)^2},
\ee
where $R_N$ is the normal-state resistance (that we take here as $R_N\equiv
R(T=1.5 T_{BKT})$. 
Finally, to account for sample inhomogeneity, we map the spatial inhomogeneity of
the sample in a random-resistor-network problem, by associating to each
patch of stiffness $J_s^i$ a normalised resistance $\rho_i=R_i/R_N$ obtained from Eq.\ 
\pref{res} by using the corresponding local values of $T_c^i$ and $T_{BKT}^i$
computed above. The overall
sample normalised resistance $\rho=R/R_N$ is then calculated in the so-called 
effective-medium-theory (EMT) approximation\cite{sema}, where $\rho$
is the solution of the self-consistent equation
\be
\lb{semares}
\sum_i \frac{w_i (\rho-\rho_i)}{\rho+\rho_i}=0,
\ee
and $w_i$ is the occurrence probability of each resistor, i.e. of the
corresponding $J_s^i$ value, as determined by the analysis below
$T_{BKT}$. As it has been discussed in Ref.\ \cite{caprara_prb11}, the EMT
approach turns out to be in excellent agreement with the exact numerical
results for a network of resistors undergoing a metal-superconductor
transition, even in the presence of SC fluctuations.  We can then employ
Eq.\ \pref{semares} to compute $R/R_N$ of our samples, by using the
probability distribution of width $\delta$ known from the analysis of $\l(T)$, and by treating $A$ and $b$ as free
parameters. The resulting fits are in excellent agreement with the
experimental data (Fig. 1c). Moreover, considering that the
interpolation formula \pref{corr} between the BKT and GL fluctuation regime
is necessarily an approximation, the obtained values of $b$ are in very
good agreement with the theoretical estimate \pref{btheo} (Table
I). Thus, our analysis above $T_c$ not only provides us with a remarkable
example of interpolation between the GL and BKT fluctuation regimes, but it
also demonstrates the validity of the values of $\mu$ obtained from $\l(T)$. Finally, we would like to
stress that $b$ cannot be used completely as a free parameter while fitting
the $\rho(T)$. Attempting to fit the BKT
fluctuation regime at $T$$\gg$$T_{BKT}$ (as proposed
in the literature [\onlinecite{triscone_nature08,iwasa_natmat10}]),
results in unphysical $b$ values with respect to relation \pref{btheo}.

\begin{table}[t]
\begin{center}
\caption{Sheet resistance ($R_s$), Magnetic penetration depth ($\l(T\ra 0)$), $T_{BKT}$, BCS
  transition temperature $T_c$, along with the best fit parameters (see
  text) obtained from the BKT  fits of $\l^{-2}(T)$ below
  $T_{BKT}$ and of $R(T)$ above $T_{BKT}$ for NbN thin
  films of different thickness $d$. The temperature $T_c$ is obtained by
  the extrapolation of the BCS fit of $\l^{-2}$ well below
  $T_{BKT}$.}
\label{t-table}

\begin{tabular}{|c|c|c|c|c|c|c|c|c|c|}
\hline 
$d$ & $R_s$ & $\l(0)$ & $T_{BKT}$ & $T_c$ &\multicolumn{3}{|c|}{Fit of $\l^{-2}(T)$} &
\multicolumn{2}{|c|}{Fit of $R(T)$} \\
\cline{6-10}
(nm) & ($k$$\Omega$) & (nm) & (K) & (K) & $\mu/J_s$ & $\d$ & $b_{theo}$ & $A$ & $b$ \\
\hline
3 & 1.2 & 582 & 7.77 & 8.3 & 1.19 & 0.02 & 0.108 & 1.35 & 0.108 \\
6 & 0.44 & 438 & 10.85 & 11.4 & 0.61 & 0.005 & 0.048 & 1.3 & 0.067 \\
12 & 0.19 & 403 & 12.46 & 12.8 & 0.46 & 0.0015 & 0.027 & 1.21 & 0.039 \\
18 & 0.1 & 383 & --& 13.37 & -- & -- & -- & -- & -- \\
\hline 
\end{tabular}
\end{center}
\end{table}

\begin{figure}[ttt]
\includegraphics[scale=0.5,clip=]{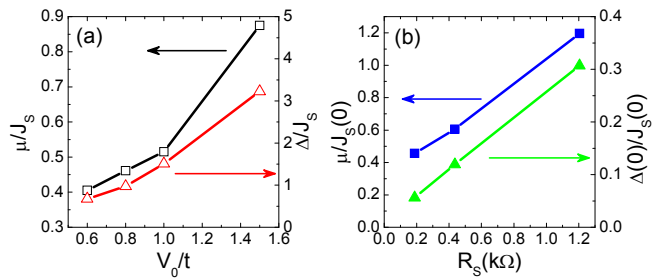}
\caption{(Color online) (a) Numerical results for the disorder dependence of $\mu/J_s$ and
  $\Delta/J_s$ as a function of disorder for the 
attractive Hubbard model. (b) Experimental values for the same ratios in
our NbN films, plotted as a function of the normal-state sheet resistance $R_S$.}
\label{fig-ratio}
\end{figure}

Once established the robustness of our estimate of $\mu$, we discuss now the
values reported in Table I, and their thickness
dependence. We first notice that the values of $\mu$ obtained by our fit are of the order of magnitude of the
standard expectation for a BCS superconductor. Indeed, in this case one
usually\cite{pra} estimates the $\mu$ as the loss in condensation energy
within a vortex core of size of the order of the coherence length
$\xi_0$, 
\be
\lb{econd}
\mu=\pi \xi_0^2 \epsilon_{cond}
\ee
where $\e_{cond}$ is the condensation-energy density.  In the clean case Eq.\ \pref{econd}
can be expressed in terms of $J_s$ by means of the BCS relations for 
$\e_{cond}$ and $\xi_0$. Indeed, since
$\epsilon_{cond}=N(0)\Delta^2/2$, where $N(0)$ is the density of states at the
Fermi level and $\Delta$ is the BCS gap, and 
$\xi_0=\xi_{BCS}=\hbar v_F/\pi \Delta$, where $v_F$ is the Fermi velocity, and  assuming
that $n_s=n$ at $T=0$, where $n=2N(0)v_F^2 m/3$, one has
\be
\lb{mu}
\mu_{BCS}= \frac{\pi \hbar^2 n_s}{4 m} \frac{3}{\pi^2}=\pi J_s \frac{3}{\pi^2}
\simeq 0.95 J_s,
\ee
so that it is quite smaller than in the $XY$-model case \pref{muxy}. While
the exact determination of $\mu$ depends on small numerical factors that
can slightly affect the above estimate, the main ingredient that we should
still account for is the effect of disorder, that can alter the relation
between $\epsilon_{cond}$, $\Delta$ and $J_s$ and explain the variations
observed experimentally. To properly account for it we computed explicitly
both $\mu$ and $J_s$ within the attractive two-dimensional Hubbard model
with local disorder:
\begin{equation}
\lb{defh}
H=-t \sum_{\langle ij \rangle  \sigma}c^\dagger_{i\sigma}c_{j\sigma}+h.c. -
|U|\sum_{i}n_{i\uparrow}n_{i\downarrow} +\sum_{i\sigma}V_i
n_{i\sigma}, 
\end{equation}
which we solve in mean field using the BdG equations\cite{degennes}.
The first sum is over nearest-neighbors pairs and we work on a
$N=N_x\times N_y$ system, with a local potential $V_i$ randomly
distributed between $0 \le V_i \le V_0$. $J_s$ is computed by the change in the ground-state energy in
the presence of a constant vector potential\cite{benfatto_bdg}, while 
$\mu$ is computed  by means of Eq.\ \pref{econd}, by determing
both $\epsilon_{cond}$ and $\xi$ in the
presence of disorder\cite{seibold} at doping $n=0.87$ and coupling $U/t=1$. 
The resulting value of $\mu/J_s$ at $T=0$ is reported in
Fig.\ 2a: It is of order of the BCS estimate and it shows a steady increase as 
disorder increases, in agreement with the experimental results, shown in
Fig. 2b, where we take the normal state sheet resistance $R_s$
as a measure of disorder as the film thickness
decreases. This behavior can be understood as a consequence of the 
increasing separation with disorder between the 
energy scales associated respectively to the $\Delta$, which controls $\epsilon_{cond}$, and $J_s$, as it is shown
by the ratio $\Delta/J_s$ that we report in the two panels of Fig. 2
for comparison. Notice that even though we used a weaker coupling $U/t=1$ as
compared to other recent studies \cite{randeria10,benfatto_bdg} this is 
still a large coupling strength as compared to our NbN samples, so that the
numerical values of $\Delta/J_s$ are larger than experimental
ones\cite{note_size}. Nonetheless, our approach already captures the
experimental trend of $\mu/J_s$ as a function of disorder,
and its correlation with the $\Delta/J_s$ behavior at large disorder. 

In summary, we have shown that to correctly identify the typical signatures
of the BKT transition in thin films of NbN we must properly account for
$\mu$ values smaller than expected within the standard
approach based on the $XY$ model\cite{bkt,bkt2}. We also
observe a steady increases of the ratio $\mu/J_s$ as the film thickness decreases. This effect can be
understood within a model for disordered superconductors, resulting from
increasing separation between the energy scales associated with $\Delta$ and $J_s$. It would be interesting to investigate if a
similar effect could be at play also in other systems, as disordered films
of InO$_x$\cite{armitage_cm10},  or high-temperature cuprate
superconductors, where a large $\mu$ value has been indirectly suggested by the analysis of
the superfluid density data\cite{benfatto_mu}.

\section{Acknowldegments}
We acknowldege Vivas Bagwe and John Jesudasan for technical support.


\begin{thebibliography}{cc}
\bibitem{note_size}
The numerical analysis of weaker couplings would require a substantial
increase of the system size. We already take much larger ($N=600$
sites) than analogous studies\cite{randeria10,benfatto_bdg} of disordered superconductors.
\end{thebibliography}
\end{document}